\documentclass[twocolumn]{aa}
\usepackage{graphicx}
\usepackage{txfonts}
\begin{document}
   \title{Non-LTE modelling of the He\,I\,10830\,{\AA} line\\ in 
early-type main sequence stars}


   \author{N. Przybilla}


   \institute{Dr. Remeis-Sternwarte Bamberg, 
Universit\"at Erlangen-N\"urnberg, 
Sternwartstrasse 7, D-96049 Bamberg,
Germany\\
\email{przybilla@sternwarte.uni-erlangen.de}
}

   \date{Received; accepted}

   \abstract{
The near-IR \ion{He}{i}\,10\,830\,{\AA} transition is a highly sensitive diagnostic for
non-LTE effects in astrophysical plasmas. So far, non-LTE line-formation
computations have failed to quantitatively reproduce observations of this line
in the entire range of early-A to late-O main sequence stars. 
It is shown that the non-LTE modelling
was insufficient, for the most part, either because of
inaccurate photoionization cross-sections for the $2s$\,$^3S$ state or
because of neglecting line blocking.
New calculations based on state-of-the-art atomic data give excellent
agreement with observation for the \ion{He}{i}\,10\,830\,{\AA} feature, while 
profiles of the \ion{He}{i} lines in the visual are retained.
   \keywords{line: formation -- infrared: stars -- stars: early-types}
   }
   \maketitle


\section{Introduction}
The 10\,830\,{\AA} transition ($2s$\,$^3S$--$2p$\,$^3P^{\circ}$) in neutral
helium is very important for constraining non-LTE effects in the second
most abundant element. It involves the metastable state of the triplet spin system
and is consequently prone to non-LTE effects, because the triplets are only 
loosely coupled via collisions to the singlet system.
In addition, its location in the Rayleigh-Jeans tail of the energy distribution
of hot stars gives rise to amplified non-LTE effects in this class
of objects. 

Since detection of this line in the solar disk spectrum (Babcock \&
Babcock~\cite{BaBa34}), it has been observed in most kinds of stars
throughout the Hertzsprung-Russell diagram (HRD). It is widely used as an indicator 
of chromospheric activity in the Sun and other cool stars (e.g. Zirin~\cite{Zirin82}). 
Towards the hot end of the HRD, this feature is observed as a 
prominent emission line, as in O-stars (e.g. Andrillat \&
Vreux~\cite{AnVr79}) and in Wolf-Rayet stars (e.g. Howarth \&
Schmutz~\cite{HoSch92}). The \ion{He}{i} 10\,830\,{\AA} transition is also 
discussed in a galactic and cosmological context,
ranging from local gaseous nebulae to Seyfert
galaxies and QSOs (e.g. LeVan et al.~\cite{LeVanetal84}). The plasma
conditions in chromospheres, stellar winds, and gaseous nebulae lead to rather
complex physics of the non-LTE line formation (e.g. Avrett et
al.~\cite{Avrettetal94}; Peimbert \& Torres-Peimbert~\cite{PeTo87}).
Therefore it is impossible to separate inaccuracies of input atomic data
from shortcomings in the plasma physics model. Consequently it is mandatory
to study the transition in a well-understood environment,
such as the stellar atmospheres of early-type stars with
negligible mass-loss. Main sequence stars of spectral types early-A to late-O
are excellent testbeds. However, non-LTE line-formation computations
for stars in this region of the HRD (Auer \& Mihalas~\cite{AuMi72},\cite{AuMi73}~-- 
AM72/73; Dufton \& McKeith~\cite{DuMK80} -- DMK80; Leone et
al.~\cite{Leoneetal95} -- LLP95) have at best reproduced observations
(Meisel et al.~\cite{Meiseletal82}; Lennon \& Dufton~\cite{LeDu89} -- LD89;
LLP95) only qualitatively so far. 
The present study resolves the issue.
Naturally, the findings will have broader implications than for this objective alone.


\section{Model calculations}
\begin{figure}
\centering
\includegraphics[width=.485\textwidth]{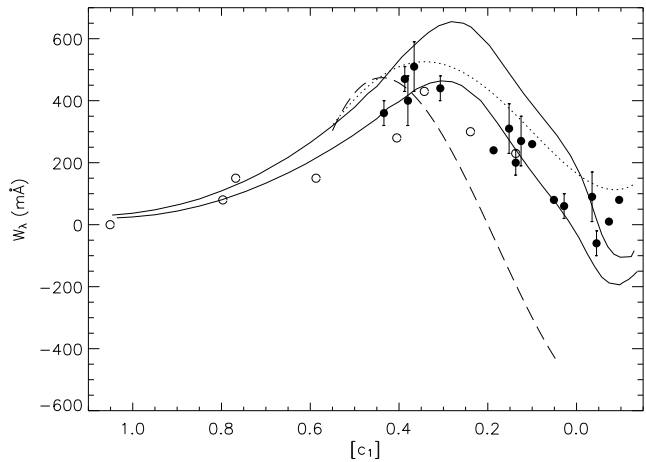}\\[-1mm]
\caption{Comparison of observed equivalent widths for
\ion{He}{i}\,10830\,{\AA} (Lennon \& Dufton~\cite{LeDu89}: filled circles;
Leone et al.~\cite{Leoneetal95}: open circles) with model predictions: Auer
\& Mihalas~(\cite{AuMi73}, dotted line), Dufton \& McKeith~(\cite{DuMK80}, 
dashed line), and present computations (full lines)
for $\xi$\,$=$\,0 (lower) and 8\,km\,s$^{-1}$ (upper curve). The abscissa is
the reddening-free $[ c_1 ]$ index. Measurements of LD89 without 
error bars have uncertainties $>$80\,m{\AA}. Uncertainties for the 
fast rotators of the LLP95 sample can be well above 100\,m{\AA}.
}
\label{ews}
\end{figure}
Two \ion{He}{i} model atoms are used in the present work: {\sc i)} the
`standard' model of Husfeld et al.~(\cite{Husfeldetal89}), which is a
slightly improved version of the AM73 model atom, later
updated with collisional excitation data of Berrington \&
Kingston~(\cite{BeKi87}), and {\sc ii)} an extended `new' version of the
Husfeld et al. model.
Considering the term structure of the new model atom, the difference is that now all
states up to principal quantum number $n$\,$=$\,5 are treated individually
(previously up to $n$\,$=$\,4), with the remainder grouped into combined levels
for each $n$ in the singlet and triplet spin systems.
Oscillator strengths are adopted from the NIST database
(http://physics.nist.gov/cgi-bin/AtData/main\_asd), from Fernley et
al.~(\cite{Fernleyetal87}, FTS87), or from the Coulomb approximation, in this order of
preference. Photoionization cross-sections of FTS87
are employed, with missing data assumed to be hydrogenic (Mihalas~\cite{Mihalas78}, p. 99). 
Effective collision strengths for 
excitation by electron collisions are adopted from {\em ab-initio} 
computations of Bray et al.~(\cite{Brayetal00}) and Sawey \& Berrington~(\cite{SaBe93}).
Additional transitions are treated according to Mihalas \&
Stone~(\cite{MiSt68}), and for the remainder of the optically forbidden
transitions, the semiempirical Allen formula (Allen~\cite{Allen73}) is applied.
Here collision strengths $\Omega$ varying between 0.001 to 1000 are
assumed, based on trends derived from examination of the detailed
computations of Bray et al. and Sawey \& Berrington. The lowest $\Omega$
occur for transitions between the ground state and the levels of highest
excitation energy, while detailed balance is established between high-lying
energy levels due to the large collision rates. Data from Bell et
al.~(\cite{Belletal83}) have been
adopted for electron impact ionization of the ground state, along with the
Seaton~(\cite{Seaton62}) approximation for the other individual levels with
threshold photoionization cross-sections from FTS87
where available, or from the hydrogenic
approximation. Collisional ionization of the packed levels is accounted for
according to Mihalas \& Stone~(\cite{MiSt68}). Finally, the theory of
Barnard et al.~(\cite{Barnardetal69}), which is described further
in AM72, is utilised for a realistic description of line broadening.
The model for \ion{He}{ii} considers all levels/transitions up to $n$\,$=$\,20.
Its details are of no further importance for the present investigation.

The non-LTE line-formation calculations are performed 
in a hybrid approach. Based on hydrostatic, plane-parallel,
line-blanketed LTE models calculated with the {\sc Atlas9} code (Kurucz~1993),
the non-LTE line formation is solved using {\sc Detail} and {\sc Surface}
(Giddings~1981; Butler \& Giddings~1985). The coupled radiative transfer and
statistical equilibrium equations are solved employing an accelerated
lambda iteration scheme that uses the treatment of Rybicki \&
Hummer~(\cite{RyHu91}). Line-blocking
is realised by considering Kurucz' opacity distribution functions. 
Note that microturbulence is explicitly accounted for in the solution of
the statistical equilibrium and radiative transfer by inclusion of an
additional term in the width of the (depth-dependent) Doppler profiles
adopted at that stage of the calculations. The grid comprises models for
effective temperature $T_{\rm eff}$ ranging from 10\,000 to 35\,000\,K at
a single surface gravity value of
$\log g$\,$=$\,4.0\,(cgs), which is adequate for early-type dwarfs with
negligible mass-loss. Solar composition is assumed, with the microturbulent
velocity~$\xi$ fixed to two values of 0 and 8\,km\,s$^{-1}$. 
Theoretical reddening-free Str\"omgren $[c_1]$-indices (where 
$[c_1]$\,$=$\,$c_1-0.2(b-y)$) are determined for the {\sc Atlas9} model atmospheres 
using the suite of programs for synthetic photometry by Kurucz. The
$[c_1]$-index parameterises the effective temperature for the comparison
with observation.

The computed \ion{He}{i/ii} lines at visual wavelengths in several of the hotter models
(27\,500\,$\leq$\,$T_{\rm eff}$\,$\leq$\,35\,000\,K)  have been compared
with state-of-the-art non-LTE model atmosphere calculations of 
Lanz \& Hubeny~(\cite{LaHu03}, LH03). 
Excellent agreement is found for the \ion{He}{i} triplets and \ion{He}{ii} lines 
(and also the hydrogen Balmer lines)
except for the utmost line cores, indicating that non-LTE effects on the 
atmospheric structure are small at line-formation depths 
and confirming the validity of the hybrid approach. The present computations
predict stronger \ion{He}{i} singlets than LH03, with the discrepancy
increasing with decreasing surface gravity. Note that the present modelling
gives better agreement with the observed visual spectrum of the standard star 
\object{$\tau$\,Sco}. The \ion{He}{i} 10\,830\,{\AA} feature is not included
in the spectrum synthesis of LH03.


\section{Comparison with observation}
New non-LTE line-formation computations have to prove their worth by
improving the model predictions when confronted with observation.
Spectroscopic investigations of the \ion{He}{i} 10\,830\,{\AA} transition in 
early-type main sequence stars are scarce, because of the reduced flux of
this type of star, contamination of the spectra by telluric absorption
at these wavelengths, and low sensitivity of standard CCDs in the $J$ band.
Only two sources of measurements of equivalent widths $W_{\lambda}$
from observations with
modern instrumentation are available from the literature: a sample of 16
slow-rotating ($v \sin i$\,$\lesssim$\,30\,km\,s$^{-1}$) B6--O9
stars from LD89 and 8 normal A1 to B2 dwarfs and subgiants from LLP95 (their
chemically peculiar objects are omitted in the following), almost all of them
being fast rotators ($v \sin i$\,$\gtrsim$\,130\,km\,s$^{-1}$). Both
samples have \object{$\gamma$\,Peg} in common, for which satisfactory
agreement of the equivalent width is found.

\begin{figure}
\centering
\includegraphics[width=.485\textwidth]{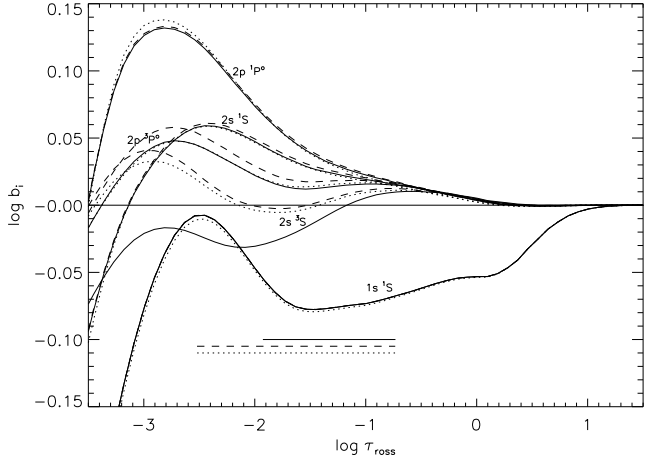}
\caption{Comparison of departure coefficients $b_i$ as a function of
Rosseland optical depth $\tau_{\rm ross}$ for the $n$\,$=$\,1 and 2
levels of \ion{He}{i} from computations using the standard (dotted lines)
and the present (full lines) model atom, and computations for the new model
atom with the photoionization cross-section for the $2s$\,$^3S$ state replaced by
that used in the old one (dashed lines). Good agreement is found except for
the $2s$\,$^3S$ state, the lower level of the \ion{He}{i}\,10\,830\,{\AA}
transition. Line-formation depths for this feature are indicated. The computations are for
a stellar atmospheric model with $T_{\rm eff}$\,$=$\,30\,000\,K, $\log
g$\,$=$\,4.0\,(cgs), zero microturbulence, solar metallicity, and solar 
helium abundance.
}
\label{depB}
\end{figure}

A comparison of equivalent widths from the present non-LTE computations with 
observation is made in Fig.~\ref{ews}, where previous model predictions 
are also summarised, employing the same $[c_1]$ calibrations as in LD89.
As already found by LD89, the calculations of AM73 systematically predict too strong
equivalent widths for \ion{He}{i} 10\,830\,{\AA}, while those of DMK80
indicate that this line goes from absorption into (strong) emission
near spectral type B2. Their calculations do not cover effective
temperatures below 15\,000\,K. Hybrid non-LTE computations based on
line-blanketed {\sc Atlas9} atmospheres by LLP95 (not shown)
find good agreement in the cooler stars, and
a qualitatively similar behaviour as DMK80 predicting slightly
less pronounced emission for spectral types earlier than B2.
Observation, on the other hand, indicates a maximum of
equivalent widths around spectral type B2, levelling off towards the late-O
stars and eventually leading to small emission, as is the case in
$\tau$\,Sco. Consequently, none of the non-LTE line-formation
computations for \ion{He}{i} 10\,830\,{\AA} from the literature
reproduce the trend in observed equivalent width of dwarf stars over the whole
range of spectral types from early-A to late-O. 

The new calculations succeed here, finding excellent agreement with
observations, as can be seen in Fig.~\ref{ews}. Note that the principal error in 
equivalent width measurements
comes from continuum placement, such that the uncertainties can be expected
to be largest in the fast rotators of the LLP95 sample. 
A correct location of the maximum of line strength is indicated and much weaker 
emission than in previous cases, reaching a maximum at $T_{\rm
eff}$\,$\simeq$\,31\,000\,K. The emission is reduced for higher microturbulent 
velocities and levels off towards higher temperatures as helium becomes ionized. 
The observational trends imply near-solar helium abundances and low
microturbulent velocities in the cooler stars and enhanced helium and/or
higher microturbulences for the earlier types (the more massive objects).
This agrees well with the findings from analyses 
of early B-stars in the classical visual wavelength range 
(e.g. Lyubimkov et al.~\cite{Lyubimkovetal04})
and predictions from recent stellar evolution calculations (e.g. Heger \&
Langer~\cite{HeLa00}; Meynet \& Maeder~\cite{MeMa00}). The results are
reassuring; however, the reasons for the improvements in the predictions 
need to be understood quantitatively. This is investigated in the following.


\section{Discussion}
\begin{figure}
\centering
\includegraphics[width=.485\textwidth]{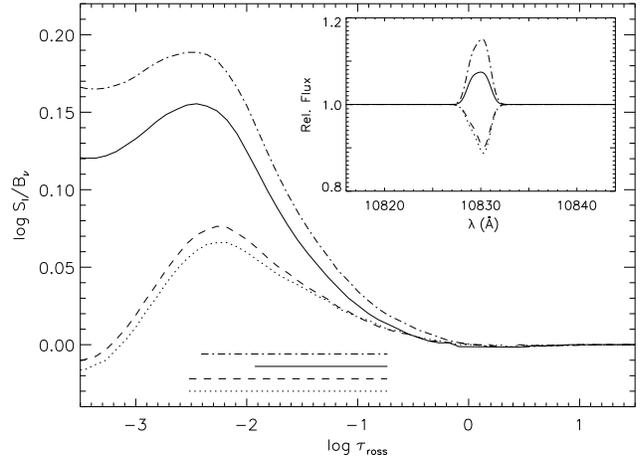}
\caption{Ratio of the line source function $S_l$ to Planck
function $B_{\nu}\,(T)$ at the centre of the 10\,830.34\,{\AA} fine-structure 
component of the transition, for the same atmospheric parameters as in Fig.~\ref{depB}.
The same line designations as in Fig.~\ref{depB} apply. In addition, results from a 
computation using the new model atom, but neglecting line blocking (dashed-dotted 
lines), are also shown. In the inset the resulting line profiles are compared after 
convolution with a Gaussian of FWHM of 20\,km\,s$^{-1}$. 
Use of the improved photoionization cross-sections turns the absorption into
emission, while neglect of line blocking effects gives excess emission.
}
\label{soverb}
\end{figure}
A comparison of the predictions from several non-LTE calculations is 
made for atmospheric parameters typical of an early B-type main sequence
star. The computations are based on boththe standard and the new model atom, and
a version of the latter where the photoionization cross-section for the
$2s$\,$^3S$ state is replaced by the one used in the old model atom.
Departure coefficients $b_i$\,$=$\,$n_i/n_i^*$ (the $n_i$ and 
$n_i^*$ being the non-LTE and LTE populations of level $i$, respectively)
for the $n$\,$=$\,1 and 2 levels of \ion{He}{i} are displayed in
Fig.~\ref{depB}. Excellent agreement is found, except for the $2s$\,$^3S$
state (and minor differences for the $2p$\,$^3P^{\circ}$ level).
The new model atom predicts level populations by up to $\sim$10\% lower at
line-formation depths than the other two models, which mutually deviate much
less. 

\begin{figure}
\centering
\includegraphics[width=.485\textwidth]{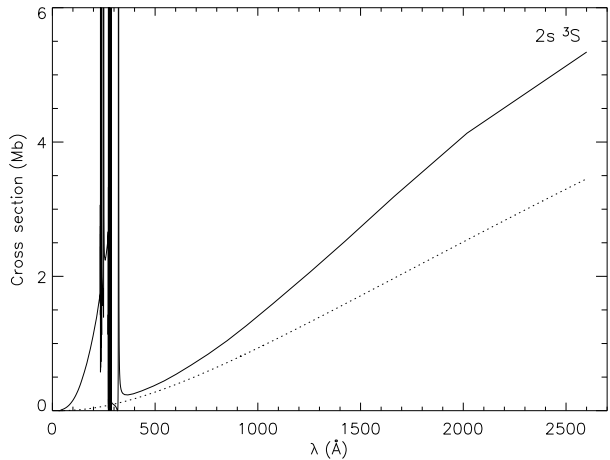}
\caption{Comparison of the photoionization cross-sections for the $2s$\,$^3S$
state of Fernley et al.~(\cite{Fernleyetal87}, full line), as used in the
present \ion{He}{i} non-LTE model atom and of Gingerich~(\cite{Gingerich64},
dotted line), as used in the AM73 and Husfeld et al.~(\cite{Husfeldetal89}) models.
The cross sections differ by more than 50\% at threshold.
}
\label{cross}
\end{figure}

Because of the high sensitivity of the line-source
function $S_l$ to variations of the departure coefficients in the Rayleigh-Jeans
limit (see e.g. Przybilla \& Butler~\cite{PrBu04}), this relatively small
difference in level populations can result in
fundamentally altered line strengths for \ion{He}{i} 10\,830\,{\AA}. While a 
ratio of $S_l/B_{\nu}$ ($B_{\nu}$ being the Planck function) slightly larger than 
unity in the old model atom (and the modified new one) leads to reduced absorption 
when compared to detailed balance predictions, the significantly larger values 
in the new model give net emission (see Fig.~\ref{soverb}).
This is primarily because of the considerable difference in the photoionization
cross-sections for the metastable state, the FTS87 data being considerably
larger, by more than 50\% at threshold, see Fig.~\ref{cross}. This weakness in the
AM73 work has already been pointed out by DMK80, but see also LD89. Note
that the detailed resonance structure of the FTS87 cross-sections, 
which are constrained to a narrow region at short wavelengths, are not
accounted for in the current computations.

The importance of {\em line-blanketed} model atmospheres on the formation 
of \ion{He}{i} 10\,830\,{\AA} (despite general systematic effects on the 
atmospheric structure through a steepened temperature gradient) is less 
pronounced than indicated by LD89, as LLP95 and the present study
use the same model atmospheres. In fact, the difference between the DMK80
and LLP95 studies, on the one hand, and ours on the other, is the
allowance for {\em line blocking}.
We therefore computed an additional model neglecting line blocking as in
the older work.
Much larger emission results from this, enhanced by a factor $\sim$2 in the
case discussed in Fig.~\ref{depB}. This is facilitated by significantly
larger ionization of \ion{He}{i} by the unblocked UV-radiation, which gives
(markedly) lower non-LTE populations for the ground state and the $n$\,$=$\,2 levels.
In particular, the $2s$\,$^3S$ state experiences stronger depopulation
relative to $2p$\,$^3P^{\circ}$, giving rise to the stronger emission in
\ion{He}{i} 10\,830\,{\AA} in the hotter stars. The agreement of the LLP95
computations with observation for the cooler stars is also explained in this
picture, as photoionization of \ion{He}{i} is inefficient there. Note that
the profiles of the \ion{He}{i} lines in the visual 
are largely insensitive to the effects of line blocking.


\section{Conclusions}
The present study resolves long-standing discrepancies between non-LTE
computations and observations for the \ion{He}{i} 10\,830\,{\AA} transition
in early-type main sequence stars. Excellent quantitative agreement is
indicated now on the level of accuracy of equivalent width 
measurements from the literature. Earlier non-LTE analyses suffered from
inaccurate photoionization cross-sections
for the $2s$\,$^3S$ state and/or neglected line blocking.
The profiles of the \ion{He}{i} lines in the visual are largely insensitive to these
details. 

Further efforts for constraining non-LTE effects for helium are
nonetheless required, as the quantitative interpretation of a wide range of 
astronomical objects relies on accurate line-formation computations for this
element. These should consider high-S/N and high-resolution
observations including the complete near-IR range as facilitated by modern
telescopes and instrumentation. Line-profile information for all available
indicators will allow stringent constraints to be put on the \ion{He}{i} model
atom.


\begin{acknowledgements}
I wish to thank K. Butler for valuable suggestions and careful reading of the manuscript. 
\end{acknowledgements}



\begin{thebibliography}{}

\bibitem[1973]{Allen73}
 Allen, C.W.~1973, Astrophysical Quantities, 3rd edition (London: Athlone Press)

\bibitem[1979]{AnVr79}
 Andrillat, Y., \& Vreux, J.M. 1979, \aap, 76, 221

\bibitem[1972]{AuMi72}
 Auer, L.H., \& Mihalas, D.~1972, \apjs, 24, 193 (AM72)

\bibitem[1973]{AuMi73}
 Auer, L.H., \& Mihalas, D.~1973, \apjs, 25, 433 (AM73)

\bibitem[1994]{Avrettetal94}
 Avrett, E.H., Fontenla, J.M., \& Loeser, R.~1994, in Infrared Solar
Physics, ed. D.M. Rabin, J.T. Jefferies, \& C. Lindsey (Dordrecht: Kluwer), 35

\bibitem[1934]{BaBa34}
 Babcock, H.D., \& Babcock, H.W.~1934, \pasp, 46, 132

\bibitem[1969]{Barnardetal69}
 Barnard, A.J., Cooper, J., \& Shamey, L.J.~1969, \aap, 1, 28

\bibitem[1983]{Belletal83}
 Bell, K.L., Gilbody, H.B., Hughes, J.G., Kingston, A.E., \& Smith,
F.J.~1983, J. Phys. Chem. Ref. Data, 12, 891 

\bibitem[1987]{BeKi87}
 Berrington, K.A., \& Kingston, A.E.~1987, J. Phys. B, 20, 6631

\bibitem[2000]{Brayetal00}
 Bray, I., Burgess, A., Fursa, D.V., \& Tully, J.A.~2000, \aaps, 146, 481

\bibitem[1985]{BuGi85}
 Butler, K., \& Giddings, J.R.~1985, in Newsletter on Analysis of
 Astronomical Spectra, No.\,9 (London: Univ. London)

\bibitem[1980]{DuMK80}
 Dufton, P.L., \& McKeith, C.D.~1980, \aap, 81, 8 (DMK80)

\bibitem[1987]{Fernleyetal87}
 Fernley, J.A., Taylor K.T., \& Seaton M.J.~1987, J. Phys. B, 20, 6457 (FTS87)

\bibitem[1981]{Giddings81}
 Giddings, J.R. 1981, Ph.\,D. thesis, Univ. London

\bibitem[1964]{Gingerich64}
 Gingerich, O. 1964, in Proc. 1st Harvard-Smithsonian Conf. Stellar
 Atmospheres, SAO Spec. Rep. 167 (Cambridge: SAO), 17

\bibitem[2000]{HeLa00}
 Heger, A., \& Langer, N.~2000, \apj, 544, 1016

\bibitem[1992]{HoSch92}
 Howarth, I.D., \& Schmutz, W. 1992, \aap, 261, 503

\bibitem[1989]{Husfeldetal89}
 Husfeld, D., Butler, K., Heber, U., \& Drilling, J.S.~1989, \aap, 222, 150

\bibitem[1993]{Kurucz93}
 Kurucz, R.L.~1993, Kurucz CD-ROM No. 13 (Cambridge, Mass.:
 Smithsonian Astrophysical Observatory)

\bibitem[2003]{LaHu03}
 Lanz, T., \& Hubeny, I.~2003, \apjs, 146, 417 (LH03)

\bibitem[1989]{LeDu89}
 Lennon, D.J., \& Dufton, P.L. 1989, \aap, 225, 439 (LD89)

\bibitem[1995]{Leoneetal95}
 Leone, F., Lanzafame, A.C., \& Pasquini, L.~1995, \aap, 293,457 (LLP95)

\bibitem[1984]{LeVanetal84}
 LeVan, P.D., Puetter, R.C., Smith, H.E., \& Rudy, R.J.~1984, \apj, 284, 23

\bibitem[2004]{Lyubimkovetal04}
 Lyubimkov, L.S., Rostopchin, S.I., \& Lambert, D.L.~2004, \mnras, 351, 745

\bibitem[1982]{Meiseletal82}
 Meisel, D.D., Saunders, B.A., Frank, Z.A., \& Packard, M.L.~1982, \apj,
263, 759

\bibitem[2000]{MeMa00}
 Meynet, G., \& Maeder, A.~2000, \aap, 361, 101

\bibitem[1978]{Mihalas78}
 Mihalas D. 1978, Stellar Atmospheres, 2nd edition (San Francisco: 
 W.~H. Freeman and Company)

\bibitem[1968]{MiSt68}
 Mihalas, D., \& Stone, M.E.~1968, \apj, 151, 293

\bibitem[1987]{PeTo87}
 Peimbert, M., \& Torres-Peimbert, S.~1987, \rmxaa, 15, 117

\bibitem[2004]{PrBu04}
 Przybilla, N., \& Butler, K. 2004, \apj, 609, 1181

\bibitem[1991]{RyHu91}
 Rybicki, G.B., \& Hummer, D.G.~1991, \aap, 245, 171

\bibitem[1993]{SaBe93}
 Sawey, P.M.J., \& Berrington, K.A.~1993, At. Data Nucl. Data Tables, 55, 81

\bibitem[1962]{Seaton62}
 Seaton, M.J. 1962, in: Atomic and Molecular Processes (New York: Academic
Press)

\bibitem[1982]{Zirin82}
 Zirin, H. 1982, \apj, 260, 655

\end{thebibliography}
\end{document}